# PRIVACY PRESERVING DATA MINING BY USING IMPLICIT FUNCTION THEOREM

Pasupuleti Rajesh[1] and Gugulothu Narsimha[2]

[1]Department of Computer Science and Engineering, VVIT College, Guntur, India

`rajesh.pleti@gmail.com`

[2]Department of Computer Science and Engineering, JNTUH University, Hyderabad, India

`narsimha06@gmail.com`

*ABSTRACT*

*Data mining has made broad significant multidisciplinary field used in vast application domains and extracts knowledge by identifying structural relationship among the objects in large data bases. Privacy preserving data mining is a new area of data mining research for providing privacy of sensitive knowledge of information extracted from data mining system to be shared by the intended persons not to everyone to access. In this paper , we proposed a new approach of privacy preserving data mining by using implicit function theorem for secure transformation of sensitive data obtained from data mining system. we proposed two way enhanced security approach. First transforming original values of sensitive data into different partial derivatives of functional values for perturbation of data. secondly generating symmetric key value by Eigen values of jacobian matrix for secure computation. we given an example of academic sensitive data converting into vector valued functions to explain about our proposed concept and presented implementation based results of new proposed of approach.*

## 1. INTRODUCTION

Data mining is considered as one of the most important frontiers in database and information systems and promising interdisciplinary developments in Information Technology[1]. Data mining has wide application domain almost in every industry[2]. Data mining has significant concern in finding the patterns, discovery of knowledge from different business application domains by applying algorithms like classification, association, clustering etc , to finding future trends in business to grow. The knowledge extraction services from tremendous volumes of data has increased greater in vast areas of domains. This type of knowledge may also consisting of sophisticated sensitive information prevents disclose to access the information by intruders. In this paper, our objective is to device a symmetric encryption scheme based on implicit function theorem that enables prescribed privacy guarantees to be proved.

### 1.1. Privacy Preserving Data Mining

Now a day's information is becoming increasing important and in fact information is a key part in decision making in an organization. We are in the world of information era. Data is the major valuable resource of any enterprise. There is a incredible amount of sensitive data produced by day-to-day business operational applications. In addition, a major utility of huge databases today is available from external sources such as market research organizations, independent surveys and quality testing labs, scientific or economic research . Studies states that the quantity of data in a certain organization twice every 5 years. Data mining and its integrated fields efficiently determine valuable sensitive knowledge patterns from large databases, is vulnerable to exploitation[3], [4]. privacy preserving data mining is an insightful and considered as one of the most importance basis in database and information systems. An attractive novel trend for data mining research is the development of techniques that integrates privacy concerns[5], [6], [7], [8]. The outsourcing of business intelligence data and computing services in acquiring knowledge sensitive relevance analysis based on privacy preserving data mining technologies are expected to amenable future trend [9], [10].

### 1.2. Vector Valued Functions

Vector valued function is a mathematical function of one or more variables whose range is a set of multidimensional vectors or infinite dimensional vectors.

$$F(x_1, x_2, \ldots, x_n) = \begin{bmatrix} f_1(x_1 \; x_2 \ldots x_n) \\ f_2(x_1 \; x_2 \ldots x_n) \\ \ldots \quad \ldots \quad \ldots \quad \ldots \\ f_m(x_1 \; x_2 \ldots x_n) \end{bmatrix}$$

Suppose $F: R^n \to R^m$ is a function from Euclidean $n$-space to Euclidean $m$-space. In vector calculus, the Jacobian matrix is the matrix of all partial derivatives of all these functions (if they exist) can be structured in an $m$-by-$n$ matrix, then the Jacobian matrix $J$ of $F$ can be represented as

$$J = \begin{bmatrix} \dfrac{\partial f_1}{\partial x_1} & \cdots & \dfrac{\partial f_1}{\partial x_n} \\ \vdots & \ddots & \vdots \\ \dfrac{\partial f_m}{\partial x_1} & \cdots & \dfrac{\partial f_m}{\partial x_n} \end{bmatrix}$$

The significance of the Jacobian matrix represents the best linear approximation to a differentiable function near a given point (the point will be consider as given input of sensitive data). In the sense, the Jacobian is the derivatives of a multivariate function [11].

International Journal of Network Security & Its Applications (IJNSA), Vol.5, No.2, March 2013### 1.3. Implicit Function Theorem

Implicit function theorem is a tool which allows relations to be converted into functions. It does this by representing the relation as the graph of a function. The conditions guaranteeing that we can solve form of the variables in terms of p variables along with a formula for computing derivatives are given by the implicit function theorem [11]. The pairs of x and y which satisfy the first relationship y=f(x) will also satisfy the second relation $g(x,y) = y - f(x) = 0$.

The motive for calling it an "implicit" function is that it does not say absolute that y depends on x, but it is a function as the variant of x can vary y as well, in order to maintain the "equals zero" relation. An implicit function of x and y is simply any association that takes the form $g(x,y) = 0$. Any explicit function can be changed into an implicit function using the trick above, just setting $g(x,y) = y - f(x) = 0$. In theory, any implicit function could be converted into an explicit function by solving for y in terms of x. In practice, this may be rather challenging, though it is hard to solve such type of a functions.

The implicit function theorem can be stated as the function $F: R^{n+m} \to R^m$ has continuous partials. Suppose $b \in R^n$ and $a \in R^m$ with $F(a,b) = 0$. The n*n matrix that corresponds to the y partials of F is invertible. Then there exist a unique function g(x) nearby 'a' such that $F(x, g(x)) = 0$ [11]. Solving a system of m equations in n unknowns is equivalent to finding the zeros of a vector-valued function from $R^m \to R^n$ where n>m. The proposed approach of privacy preserving data mining by implicit function theorem finds the jacobian matrix of partial derivatives of such vector valued function represents the best linear approximation of function.
In this paper, we proposed a new approach that facilitate the parties to share the secure transformation for the real world sensitive information by using implicit function theorem. The paper is structured as follows. Section 2 describes about Related work. We explained our proposed approach of Privacy preserving Data mining by using Implicit Function Theorem in section 3 with an example of academic sensitive information of data. Section 4 provides the experimental result analysis. Section 5 consists of conclusion and future scope.

## 2. RELATED WORK

The problem of privacy-preserving data mining has raise to be more significant in recent years for the reason that of the increasing ability to store personal private data about users and the increasing sophistication of data mining algorithms to leverage sensitive information of data. Identification of problems associated to all aspects of privacy and security concerns in data mining are extensively growing in real time environment applications[7],[12].Two current central categories to perform data mining tasks without compromising privacy are Perturbation method and the secure computation method. However, both have a few difficulties, in the first one reduced accuracy and increased overhead for the second. By including any privacy preserving technique to data mining, the communication and computation cost will not be increased.

Majority of approaches associated to Privacy-preserving Data mining are

### 2.1. Anonymization

Sending of multiple records of same type along with original record [13].

23

International Journal of Network Security & Its Applications (IJNSA), Vol.5, No.2, March 2013## 2.2. Obfuscate

Provide confusion and randomness to sensitive values of data [5], [6], [14], [15], [16].

## 2.3. Cryptographic Hiding

Proliferation of sensitive data by encryption decryption methods and secure computation methods [17], [18], [19], [20], [21].

The extracted knowledge obtained from data mining system should not be disclose to everyone. such analysis of sensitive information has to provide privacy and protect the privacy of individual sensitive information. Privacy preserving damining framework rationally combines both features of ambiguity and certainty of original data base and covers diversity of security parameters need to be consider when outsourcing the sensitive data. In this paper we proposed a new approach of Privacy preserving Data mining by using Implicit Function Theorem to share sensitive information of knowledge in secure manner.

# 3. PROPOSED WORK

Privacy preserving data mining research having two kinds of approaches . The first one is to changing original data prior to passing through to the data mining system. so that real values of data are ambiguous(perturbation). The second approach is privacy preserving distributed data mining(secure computation). our new privacy preserving technique integrates these two approaches and provides two way enhanced security approach. In this scenario, no party knows anything excluding its own input and the results. we need some communication between the parties is required for any interesting computation. but what if the result itself violates privacy?.
We need a techniques to define and quantify privacy to ensure that privacy preserving data mining results will meet required indented purpose without disclosing sensitive information. Not revealing of sensitive information , we can ensure the secure communication by applying our new technique of "Privacy preserving Data mining by using Implicit Function Theorem". The key idea is to provide secure communication between the parties who wish to share sensitive extraction of knowledge obtained from data mining system by a symmetric key value of jacobian matrix.

By employ our new privacy preserving data mining technique using implicit function theorem, the communication cost to share the sensitive information between parties and computation time will reduced. By including privacy preserving technique in data mining using implicit function theorem, the concept of "sensitive information " cannot be known in advance to intruder because of dynamically generated secret key value of input data.

A key distinction of privacy preserving data mining by implicit function theorem is to setting not only the underlying data but also the mined results should be shared for intended persons and must remain private. This type of proposed approach will solves real-world problems.

Privacy preserving Data mining by using Implicit Function Theorem is two way method enhanced security approach includes:

1) Perturbation of sensitive knowledge of data by partial derivatives of functional values .
2) Secure computation of key by the eigen value of Jacobian matrix which satisfies the implicit function theorem

24

The generated secret key value obtained from implicit function theorem takes the input values as from sensitive kind of data. So the key value is dynamically generated from the type of input of sensitive selected values of data. The pseudo code of proposed algorithm is

**Algorithm 1.** Algorithm For Privacy Preserving Data Mining By Using Implicit Function Theorem

**Input:** Extracted data mining result or input data base to be shared
**Output**: Generate dynamic secret key to encrypt and decrypt the data.

**Step 1:** Identify the sensitive values from the input data
**Step 2:** Form the sensitive data values into vector valued functions
**Step 3:** Achieve perturbation by transforming the sensitive values of data into differentiable functional values by Implicit function theorem.
**Step 4:** Pass these values to jacobian matrix and generate eigen values
**Step 5:** Select random eigen secret key value to encrypt and decrypt the sensitive data.

The architecture of the proposed system by Privacy preserving data mining by implicit function is represented as follows let us consider an example of sensitive information of academic details of number of students in an university according to year wise [22]. our assumed vector valued data set consists of 5 variables and 3 functions. the five variables are Girls($x_1$), Boys($x_2$), Total($x_3$), Placements($x_4$), pass percentage($x_5$) and three functions are Graduation($f_1$), Post graduation($f_2$), Total($f_3$). Perturbation can be attain by transforming the original values of sensitive data into partial derivatives of the random functional values. In the example of academic sensitive data, we applied the following random functions for the year of 2011 and corresponding first order partial derivative values of sensitive data and sensitive information of university database are represented in the following way.

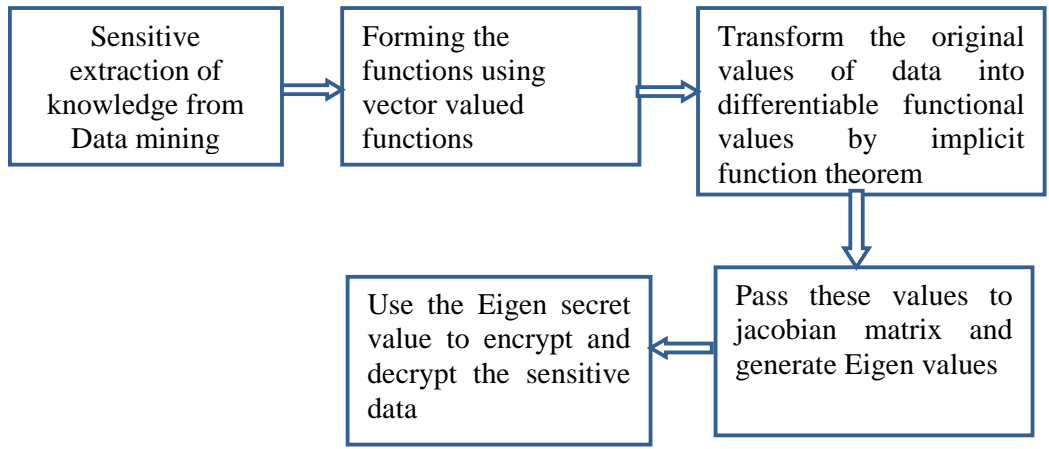

Figure 1. proposed architecture for privacy preserving Data mining

| | Year 2008(F1) | | | Year 2009(F2) | | | Year 2010(F3) | | | Year 2011(F4) | | |
|---|---|---|---|---|---|---|---|---|---|---|---|---|
| | Graduation(f1) | Post graduation(f2) | Total(f3) | Graduation(f1) | Post graduation(f2) | Total(f3) | Graduation(f1) | Post graduation(f2) | Total(f3) | Graduation(f1) | Post graduation(f2) | Total(f3) |
| Girls (x1) | 100 | 500 | 100 | 200 | 520 | 720 | 250 | 530 | 780 | 300 | 550 | 850 |
| Boys (x2) | 1300 | 100 | 1300 | 1400 | 200 | 1600 | 1450 | 250 | 1700 | 1500 | 300 | 1800 |
| Total (x3) | 1400 | 600 | 1400 | 1600 | 720 | 2320 | 1700 | 780 | 2480 | 1800 | 850 | 2650 |
| Placements (x4) | 1200 | 400 | 1200 | 1300 | 640 | 1940 | 1500 | 680 | 2180 | 1600 | 780 | 2380 |
| pass percentage (x5)(%) | 95 | 90 | 95 | 95 | 91 | 93 | 96 | 92 | 94 | 97 | 95 | 96 |

Table 1: Sensitive information of university data base

$$f_1(x_1, x_2, x_3, x_4, x_5) = x_1^2 + 2x_1x_2 + 4x_3x_4 + 5$$
$$f_2(x_1, x_2, x_3, x_4, x_5) = x_2^2 + 4x_2x_3 + 6x_1x_5 + 10$$
$$f_3(x_1, x_2, x_3, x_4, x_5) = x_3^2 + 2x_1x_4 + 5x_2x_5 + 4$$
$$\frac{\partial f_1}{\partial x_1} = 2(x_1 + x_2) = 3600, \frac{\partial f_2}{\partial x_1} = 6x_5 = 5.82,$$
$$\frac{\partial f_3}{\partial x_1} = 2x_4 = 3200.$$

$$\frac{\partial f_1}{\partial x_1} = 2x_1 = 600, \quad \frac{\partial f_2}{\partial x_2} = 2(x_2 + 2x_3) = 10,200,$$

$$\frac{\partial f_3}{\partial x_2} = 5x_5 = 485.$$

$$\frac{\partial f_1}{\partial x_3} = 4x_4 = 6400, \quad \frac{\partial f_2}{\partial x_3} = 4x_2 = 6000, \quad \frac{\partial f_3}{\partial x_3} = 2x_3 = 3600$$

Perturbation is achieved by transforming the original values into functional values of the partial derivatives. Pass these first order partial derivatives as the values in matrix. Then by implicit function theorem, the derivative of the function F (The Jacobian matrix) represents the best linear approximation to a differentiable function near a given input points of sensitive information. Compute the Eigen values of this 3×3 Jacobin matrix are 10610, -810, 7600. Choose one of the Eigen value as a secrete key to encrypt and decrypt the sensitive original information of data. In this way, we provided two way enhanced security approach consisting of perturbation and secure computation for sharing of sensitive information of data among parties.

## 4. RESULT ANALYSIS

In this section, we assess the performance of the proposed approach privacy preserving data mining by implicit function theorem. We exploit web server XAMPP 1.8.0 in order to develop the project of proposed method. XAMPP server contains features like PHP, MYSQL, Tomcat, Apache, PhPMyAdmin. We experimented with the sensitive information of academic details of university data according to year of 2011. The meaning $f_1(x_1, x_2, x_3, x_4, x_5)$ is the sensitive information of number of girls, boys total, placement details and pass percentage of graduate students in university. First we transform the original sensitive data into the partial derivatives of the random functional values by PHP code and store the corresponding perturbation values in excel format. Then export MS excel data to MySQL using the XAMPP phpmyadmin control panel.

MySQL is the trendy open source database and usually used in combination with the PHP scripting language to instruct website operations. Then retrieve the perturbation values from MYSQL data and compute eigen values by pass perturbation values to jacobian matrix with PHP code. Employ one of the eigen value of jacobian matrix as a symmetric key to encrypt and decrypt the sensitive information between parties.

Privacy preserving data mining by implicit function theorem is a proliferation, who wish share the sensitive information among parties by incorporating three components. First, dynamically generated secret key based on the type of input of sensitive selected values of data. Second, providing more uncertainty to attacker by transforming the original values of sensitive data into perturbation values. Third, secure computation of sensitive information with the generated secret key. The results shows the trade of between the privacy data quality and utility of data. In our experiment, the instant of time taken for perturbation and secure computation of sensitive values of data provides a linear approximation time, which represents best fit for the model.

If the intruder tries to find out the communication path or want to modify the message contents, he cannot achieve, its hardening that because of two way enhanced security of approach and dynamically generated secret key value based on sensitive input data. The computational cost of encryption and decryption is linear and reasonable as it is varies along with the amount of data to be send. The following graph shows time complexity based analysis to encrypt and decrypt sensitive information of data in seconds for various kinds of data sizes in KB's.

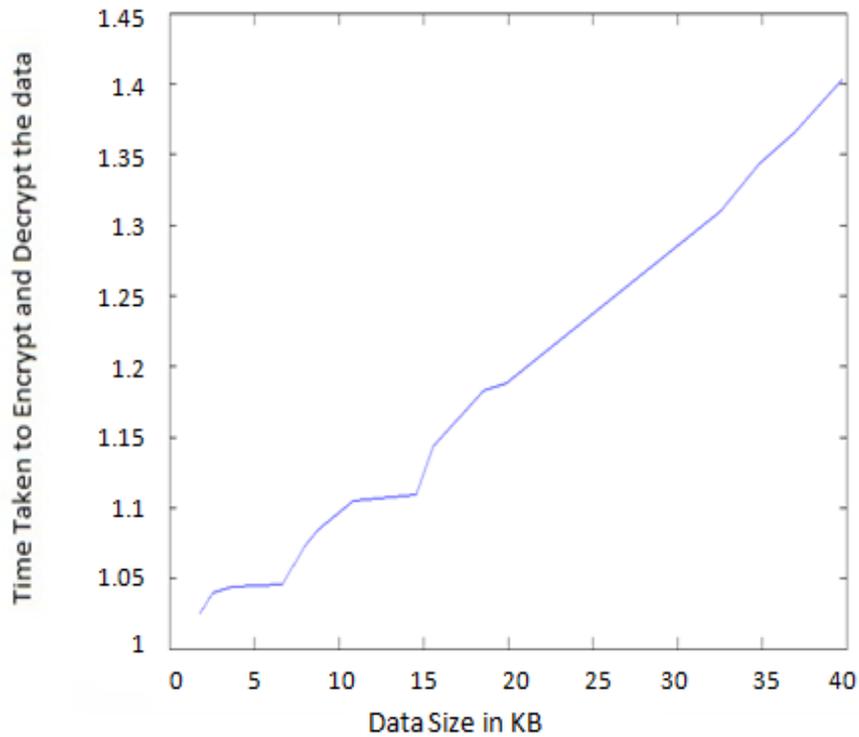

Figure 2. Time execution analysis for various data sizes

The output of the proposed method executed by XAMPP web server is showed in the following figure.

```
Assumed matrix values are:3600 ,5.82, 3200 ,600, 10200 ,4.85, 6400 , 6000, 3600
The generated three eigen values are as below:
10610
-810
7600
The secret key is:
Key: 10610

The plain text which we want to encrypt is:
An institution contains 2000 students in that there are 600 female
students, 1400 male students. In that 500 female students are post
graduate students, 100 male students are graduate students and 100 female
students are graduates, 1300 male students are graduates. In that 1200
graduate students are placed in companies and 400 post graduate students
are placed in companies. Pass percentage of graduate students are 95
percentage and pass percentage of post graduate students are 90
percentage.
the cipher text which we want to decrypt is:
P~6x-ov{jdd ~~<a}peq  c>  &1}}nw>bdcuurva>pbs1wncvkpds1chwv{ de1qrf2/!
6wuqc~{1cbdtylfm1qdt0{pszdqbtc0"#-!
6|qpg2meert~hq2  cu6vb}fg  eue?0U12jyqb1!.2">vbwue}vw>bdcuurva>pbs1`pcq{u0
  o  m  np~  tc<c|z1$&!01maj1wdpticf{1cbdtylfm1qdt0lns}tt6x-<a}saqxxuo,2npce1
`ypq{  dwvuvbwue}vw>bdcuurva>pbs1)}"b{css  d}ew>p~r1`}qa>audrurvsyt0yw01maj
1wdpticf{1cbdtylfm1qdt0%22ntbut~hcu{?
The decrypted plain text is:
An institution contains 2000 students in that there are 600 female
students, 1400 male students. In that 500 female students are post
graduate students, 100 male students are graduate students and 100 female
students are graduates, 1300 male students are graduates. In that 1200
graduate students are placed in companies and 400 post graduate students
are placed in companies. Pass percentage of graduate students are 95
percentage and pass percentage of post graduate students are 90
percentage.
Total Executaion Time 1.0085899829865 seconds
```

International Journal of Network Security & Its Applications (IJNSA), Vol.5, No.2, March 2013

Figure 3. Output of the proposed approach

| S. No | Characteristics | Privacy preserving by implicit function theorem | Differential privacy | Cryptography | K-anonymization |
|---|---|---|---|---|---|
| 1 | Security key | Dynamically generated based on type of input sensitive data | Fixed | Fixed/ Dynamic | Fixed |
| 2 | Interface of third parties | No | Yes | Yes | Yes |
| 3 | Accuracy | High (Two Way Enhanced approach) | High | High | Low |
| 4 | Computational Complexity | Low | High | High | Low |

Table 2 : A comparison of proposed algorithm.

The above table provides technical features of various privacy preserving techniques in data mining. A common characteristic of most of the previously studied frameworks is that patterns mined from huge volumes of data may be anonymized, otherwise transformed, altered to perturbation, encryption/decryption of sensitive data and secure multi party encryption schemas. In our proposed approach of privacy preserving data mining by implicit function theorem, we proposed two way enhanced approach includes both perturbation and encryption in a linear approximation time.

privacy preserving data mining describes to avoid information disclosure due to legitimate admission to the data. Sensitive knowledge which can be extract from databases by by means of data mining algorithms to be excluded, since such knowledge can evenly well compromise information privacy.

The proposed approach of privacy preserving data mining by implicit function theorem protects the sensitive information of a data with enhanced accuracy and without loss of information which creates a model for usability of data. The sensitive information of data can also be reconstructed.
With a amount of well exposed and expensive thefts creating both remarkable legal liability and bad publicity for the effected industries , business has rapidly developed more sophisticated in defending against such attacks, but so have the hackers.

There is a emergent requirement to protect sensitive information of industry data, members of staff information, client information across the enterprise wherever such data may reside. Until in recent times, mainly data theft take place from malicious individuals hacking into production databases.

privacy preserving data mining by implicit function theorem provides the tradeoff between privacy and utility problem by considering the privacy and algorithmic requirements at the same time.

29

Privacy Preserving Data Mining (PPDM) tackles the problem of developing precise models about summarized data without access to exact information in personal individual sensitive data information. A broadly studied perturbation-based privacy preserving data mining and cryptography schema approaches introduced more confusion to adversary to obtain sensitive information there by reducing chance of vulnerability to preserve privacy prior to data are published.

## 5. CONCLUSION AND FUTURE SCOPE

Privacy preserving of information is the stage key role, because of outstanding to progress in technology, contribution of organizational precise data and usefulness of information has enlarged immensely. National and international business corporations spends billions of dollars for protecting the senstive information. Generally, many of vulnerabilities come up with the financial servicies and government organizations. Proposed algorithm of privacy preserving data mining by implicit function theorem is two way enhanced security approach can be employed for various real time application domains. Sensitive information of privacy preserving data mining will be present in the forms of analysis information similar to Medicine - hospital cost analysis, prediction hospital cost analysis, drug side effects, automotive diagnostic expert systems genetic sequence analysis. Finance - credit assessment, fraud detection stock market prediction, Marketing/sales - sales prediction, product analysis, target mailing, identifying unusual behaviour, buying patterns, Scientific discovery, Knowledge Acquisition. In addition to that, privacy preserving data mining by implicit function theorem kind of approach will also be used in distributed data mining to protect information of privacy and applied for business data, which can be represented in the form of vetor valued functions.

### ACKNOWLEDGEMENTS

The authors would like to extend their gratitude to the anonymous reviewers and who continuously supported to bring out this technical paper. I would like to thank my advisor, Dr. A. Damodaram, Director of Academic Audit Cell, Jawaharlal Nehru Technological University Hyderabad , for his professional assistance and remarkable insights. I would like to express my gratitude to Prof. G. Narsimha for helping me take the first steps in the research area. And finally, special thanks to my mother and father for providing the moral support to me. we are pleased to acknowledge our sincere thanks to our beloved chairman sri.V. Vidya Sagar and Director prof. S.R.K. Paramahamsa for valuable cooperation.

## Author

**P.Rajesh** received the M.Tech degree in computer science and engineering (CSE) from Jawaharlal Nehru Technological University Hyderabad in 2009. He is currently pursuing Ph. D degree in the department of computer science and engineering from Jawaharlal Nehru Technological University Hyderabad and working as an assistant professor in CSE department at Vasireddy Venkatadri Institute of technology, Guntur, Andhra Pradesh. His research interests are in the area of Data mining, Information security, Privacy preserving data publishing and sharing

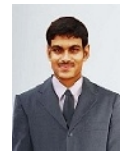

**Dr.G.Narshima** received Ph.D degree from osmania university, Hyderabad. He is having thirteen years of teaching experience and having seven years of research experience in various prestigious institutions. He is currently working as an associate professor in the department of compute**r** science and engineering from Jawaharlal Nehru Technological University Hyderabad. He has enormous research and teaching learning experience in various prestigious universities. His research interests are in databases, data privacy, Data mining, Information security, Information networks, Mobile communications, Image processing, Privacy preserving data publishing and sharing.

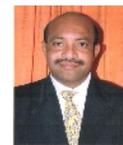